\documentclass[3p,times]{elsarticle}
\usepackage{amssymb}

\usepackage{ecrc}
\usepackage{amsthm}
\usepackage{amsmath}
\usepackage[utf8]{inputenc}
\usepackage{multirow}
\usepackage{xcolor}
\usepackage{epsfig}
\usepackage[figuresright]{rotating}
\usepackage{subfig}
\usepackage{array,booktabs}
\usepackage{hyperref}
\usepackage{graphicx}

\pdfmapfile{+txfonts.map}
\pdfmapfile{+txfonts.map}
\volume{00}
\firstpage{1}
\journalname{Physica A}
\runauth{Michele Campolieti et al.}
\jid{procs}
\jnltitlelogo{Physica A}
\CopyrightLine{2020}{Published by Elsevier Ltd.}

\input{tcilatex}

\begin{document}

\begin{frontmatter}







\title{The Distribution of Strike Size:Empirical Evidence from Europe and North America in the 19th and 20th Centuries}

 \author[label1]{Michele Campolieti\fnref{label4}}
  \author[label3]{Arturo Ramos\corref{cor1}}
 \address[label1]{Centre for Industrial Relations and Human Resources, University of Toronto, Canada}
\address[label3]{Departamento de An\'alisis Econ\'omico, Universidad de Zaragoza, Zaragoza, Spain}

\cortext[cor1]{Corresponding author\\
E-mail address: aramos@unizar.es}

\fntext[label4]{campolie@chass.utoronto.ca (M. Campolieti)}

\begin{abstract}
We study the distribution of strike size, which we measure as lost person days, for a long period in several countries of Europe and America. When we consider the full samples, the mixtures of two or three lognormals arise as very convenient models. When restricting to the upper tails, the Pareto power law becomes almost indistinguishable of the truncated lognormal.
\end{abstract}


\end{frontmatter}


\section{Introduction}

\label{intro}

Ref. \cite{Kru96} highlighted that the sizes above a threshold (however
measured) for physical, biological or economic phenomena (e.g., cities and
firms) are well described by a power law distribution. This distribution, as
well as the mechanisms that give rise to it, have also been explored in the
literature examining the size of conflicts %
\citep{Ric48,RobTur98,Ced03,ClaYouSke07,Fri15,Gon15}. More recently,
interest has also turned to the distribution of strike size %
\citep{Big05,Big18,Cam19b}, a much less violent form of conflict, but one
that can have substantial economic consequences. Interestingly, the size of
firms and unions have both been found to be described by power law
distributions (e.g., among many others, \cite{SimBon58,Axt01,Pen14}) and
strikes are embedded in these firms and unions.

The presence of a power law distribution in strike size has a few important
implications. As emphasized by \cite{Cam19b}, the total costs of strikes to
the economy will be created by a small percentage of strikes, not unlike
Pareto's rule where 20 percent of the population hold 80 percent of the
wealth. Another important implication of power law distributions is that
they may give rise to self-organizing systems, in which a system's
organization is determined by behavior of individual agents acting in
response to one another \citep{BakChe91,Kru96,Fre98}. If strikes are
self-organizing then that suggests that they are not likely to be affected
by legislation and interventions by third parties (i.e., factors outside the
``system'') and might also give rise to strike waves as they spread through
sectors or regions as more persons join them. Consequently, understanding
the distribution of strike size is quite important from the perspective of
many economic stakeholders, e.g., legislators, firms and union members. In
addition, understanding the distribution of strike size could also provide
insights into the potential models that could describe strike dynamics %
\citep{Big05,Cam19b}.


However, while the upper tail of the distribution could be fit by the power
law distribution (e.g., \cite{Big05,Big18,Cam19b}), the nature of the
distribution below the upper tail is still not understood. In addition, it
is often difficult to distinguish the power law from other alternative
distributions in the upper tail, so that alternative distributions could
also be plausible fits \citep{ClaShaNew09}. Moreover, is there a
distribution that could fit the whole range of the data, and not focus on
the upper or lower tail? Recently, papers have also begun exploring
alternatives to the power law distribution in many empirical settings and
also considering the fit below the upper tail (e.g., the size of cities %
\citep{KwoNad19,BanChiPrePueRam19,Su19,PueRamSan20}, the size of business
firms \citep{Tom20}).

We study the size distribution of strikes using micro data on strikes from a
few European countries (France\ and the Netherlands) as well as Canada and
the United States. Our data spans a long time period, covering strikes
before World War II as well as after the War. Using data from such a long
time period means that we can compare the estimates across countries during
similar periods as well different periods and so infer whether the
distribution of strikes is generated by a mechanism that is common to all
countries and all time periods. In other words, is the distribution of
strike size relatively stable across time and countries. We measure strike
size with the number of person days lost to a strike (number of workers on
strike multiplied by the duration of strike). We consider distributions such
as the lognormal and power law as well as mixtures of lognormal
distributions and truncated lognormal distributions. We obtain our estimates
using maximum likelihood estimators and to evaluate the fit of the
distributions we consider standard statistical tests as well as using
information criteria to select the most appropriate distribution. Finally,
we also discuss some of the plausible generating mechanisms for the
distributions that we consider, showing how they arise from the stationary
solutions of Fokker-Planck equations.

This paper makes the following contributions to the literature:

\begin{itemize}
\item[-] We show that the distribution of lost-person-days measure of strike
size using data in the upper tail and below it follows a mixture of two or
three lognormals.

\item[-] When restricting to the upper tail, according to the procedure of %
\cite{ClaYouSke07}, it becomes difficult to distinguish between a Pareto
power law and a (upper-tail) truncated lognormal.

\item[-] The upper tail is plausibly described by a power law, although
other models can parameterize it as good as the former.

\item[-] We formulate stochastic models that generate the lognormal mixtures
as well as the other distributions we consider.
\end{itemize}

The rest of the paper is organized as follows. The next section of the paper
presents the methodology we follow. We describe our data in Section~\ref%
{databases} and present a discussion of our findings in Section~\ref{results}%
. In Section~\ref{discussionconcl} we provide a summary of our findings and
their implications.

\section{Methods and\ Generating Mechanisms}

\label{meto}

In this section we will introduce the distributions used throughout the
paper. We let $x$ denote our measure of strike size, lost-person-days.

The first distribution we consider for lost-person-days, $x>0,$ is the
stretched exponential (STEXP) distribution
\begin{equation*}
f_{\mathrm{STEXP}}(x;\gamma ,\eta )=\frac{\gamma }{\eta }\left( \frac{x}{%
\eta }\right) ^{\gamma -1}\exp \left( -\left( \frac{x}{\eta }\right)
^{\gamma }\right)
\end{equation*}%
where $\gamma $ is a shape parameter, which is such that $0<\gamma <1$, and $%
\eta >0$ is a scale parameter. Stretched exponential distributions have
often been found to fit skewed and heavy-tailed data %
\citep{LahSor98,JiaXieLiPodZhoSta13}.

The second distribution in our study is the well-known lognormal
distribution
\begin{equation*}
f_{\mathrm{LN}}(x;\mu,\sigma )=\frac{1}{\sqrt{2\pi }\sigma x}\exp \left(-%
\frac{(\ln(x)-\mu )^{2}}{2\sigma ^{2}}\right)
\end{equation*}%
where $\mu $ is the mean of $\ln (x)$ and $\sigma >0$ is its standard
deviation according to this distribution. Lognormal distributions have been
found to fit many economic and physical phenomena with heavy tails %
\citep{Mit04}.

While lognormal distributions are skewed and can have heavy tails, they
might not be able to capture the skewness and kurtosis of some data. An
alternative that might be better able to do so is a mixture of
distributions, which include several lognormal distributions %
\citep{Ham94,McLPee00}. Mixtures of $m\geq 2$ lognormal distributions have
the density
\begin{equation*}
f_{m\mathrm{LN}}(x;\mu_{1},\sigma_{1},\ldots,
\mu_{m},\sigma_{m},p_{1},\ldots,p_{m-1}) =\sum_{j=1}^{m-1}p_{j} f_{\mathrm{LN%
}}(x;\mu_{j},\sigma_{j}) +\left(1-\sum_{j=1}^{m-1}p_{j}\right) f_{\mathrm{LN}%
}(x;\mu_{m},\sigma_{m})
\end{equation*}%
where $\mu_{1},\ldots,\mu_{m}>0$, $\sigma_{1},\ldots,\sigma_{m}>0$, and $%
0\leq p_{1},\ldots ,p_{m-1},p_{1}+\dots +p_{m-1}\leq 1$. In this paper, we
will consider 2-mixtures ($m=2$) and 3-mixtures ($m=3$), which we denote as
2LN and 3LN.

Another alternative for the distribution of strike size, which has been
considered in the earlier literature \citep{Big05,Big18,Cam19b} is the power
law distribution (or Pareto distribution):
\begin{equation*}
f_{\mathrm{P}}(x;\alpha ,x_{\mathrm{min}})=\frac{\alpha -1}{x_{\mathrm{min}}}%
\left( \frac{x}{x_{\mathrm{min}}}\right)^{-\alpha }
\end{equation*}%
where $\alpha $ is the power law exponent and $x_{\mathrm{min}},$ which is
such that $x\geq $ $x_{\mathrm{min}}>0$, is the lower bound on power law
behavior. Unlike the stretched exponential, lognormal and mixtures of
lognormal distributions we consider, which take $x>0,$ the support of the
power law distribution is $x\geq x_{\mathrm{min}}$, so it only considers the
upper tail of the data.

However, one issue that has been highlighted in the research literature is
that it can be difficult to distinguish a power law distribution with the
upper tail of other heavy tailed distributions \citep{ClaShaNew09}. This
means that there could be other distributions that are a plausible fit to
the upper tail of the data. An implication of this observation is that an
empirical analysis should also consider the fits of alternative
distributions to the upper tail \citep{ClaShaNew09}. Consequently, we also
consider a truncated lognormal
\begin{equation*}
f_{\mathrm{LNt}}(x;\mu,\sigma,x_{\mathrm{min}}) =\frac{f_{\mathrm{LN}%
}(x;\mu,\sigma)} {1-\mathrm{cdf}_{\mathrm{LN}}(x_{\mathrm{min}};\mu,\sigma)}
\end{equation*}%
where $x_{\mathrm{min}}$ is the minimum value of the variable of the range
considered ($x\in\lbrack x_{\mathrm{min}},\infty)$) and
\begin{equation*}
\mathrm{cdf}_{\mathrm{LN}}(x;\mu,\sigma) =\frac{1}{2} +\frac{1}{2}\mathrm{erf%
}\left(\frac{\ln(x)-\mu} {\sqrt{2}\sigma }\right)
\end{equation*}%
is the cumulative distribution function (CDF) of the lognormal distribution (%
$\mathrm{erf}$ is the error function associated to the standard normal
distribution). The truncated lognormal distribution has been considered as
an alternative to a power law distribution in the upper tail of wealth %
\citep{Cam18}.

We estimate all these distributions with maximum likelihood (ML) estimation.%
\footnote{%
We obtain our estimates using the command \texttt{mle} in \textsc{MATLAB$%
^{\circledR}.$} Standard errors (SE) of the ML estimators have been computed
independently using the software package \textsc{Mathematica}$^{\circledR}$
according to the indications of \cite{McCVin03} and \cite{EfrHin78}.}

We assess the fit using the log-rank/corank plots to value the fit visually.
As is well known, the tails of a power law distribution should produce a
straight line in a log-rank/corank plot. We also assess the fit of the
distributions to the data with goodness of fit tests. Namely, we use the
Kolmogorov--Smirnov (KS), Anderson--Darling (AD)\ and Cram\'{e}r--von Mises
(CM)\ tests.

The visual and statistical assessments of the models allow us to discern the
distributions that are a good fit to the data. However, the question remains
which of the distributions is the best fit or most appropriate if the tests
for fit indicate that they are plausible alternatives. Ref. \cite{Ame80}
advised that theory and economic intuition should guide model selection.
However, if they are unable to provide guidance, model selection can be
undertaken using information criteria.

We use three well-known information criteria in order to select the most
preferred model {}from those we consider. They are:

\begin{itemize}
\item[-] The Akaike Information Criterion (AIC) \cite%
{Aka74,BurAnd02,BurAnd04}, defined as
\begin{equation*}
AIC=2k-2\ln L^*
\end{equation*}
where $k$ is the number of parameters of the distribution and $\ln L^*$ is
the corresponding (maximum) log-likelihood. The minimum value of AIC
corresponds (asymptotically) to the minimum value of the Kullback--Leibler
divergence, so a model with the lowest AIC is selected from among the
competitors.

\item[-] The Bayesian or Schwarz Information Criterion (BIC) \cite%
{BurAnd02,BurAnd04,Sch78}, defined as
\begin{equation*}
BIC=k \ln(n)-2 \ln L^*
\end{equation*}
where $k$ is the number of parameters of the distribution, $n$ the sample
size and $\ln L^*$ is as before. The BIC penalizes more heavily the number
of parameters used than does the AIC. The model with the lowest BIC is
selected according to this criterion.

\item[-] The Hannan--Quinn Information Criterion (HQC) \cite%
{BurAnd02,BurAnd04,HanQui79}, defined as
\begin{equation*}
HQC=2k\ln (\ln (n))-2\ln L^{\ast }
\end{equation*}%
where $k$ is the number of parameters of the distribution, $n$ the sample
size and $\ln L^{\ast }$ is as before. The HQC implements an intermediate
penalization of the number of parameters when compared to the AIC and BIC.
The model with the lowest HQC is selected according to this criterion.
\end{itemize}

The information criteria aid us in selecting the most appropriate
distribution for the data.

A more difficult question is sorting among the mechanisms that create the
distribution of strike size. Some earlier work, \cite{Cam19b}, points to two
mechanisms that could create a power law distribution in the upper tail of
the distribution of strike size. First, the safety valve hypothesis, where
dissatisfaction and conflict builds up in a workplace reaches a critical
point and is then released as a strike. This model is like a forest fire
model, which has also been applied to study the spread of conflict
\citep{Ced03,Big05}. Second, the joint costs model, where the incidence and duration
of strikes is inversely related to the costs of striking. Both of these
mechanisms would create a power law distribution in the upper tail. However,
it is not clear whether these mechanisms would be consistent with the other
distributions, e.g., stretched exponential, the lognormal, truncated
lognormal or mixture of lognormals, that we consider. To obtain these
alternative distributions we need a more general dynamic model.

In order to achieve this task, let us denote first the natural logarithm of
the strike size variable by $y=\ln (x)$ in what follows, and then $y\in
(-\infty ,\infty )$. Then the stretched exponential becomes the
``exponential stretched exponential'':
\begin{equation*}
f_{\mathrm{ESTEXP}}(y;\gamma ,\eta )=\gamma \exp \left( -\left( \frac{%
\mathrm{e}^{y}}{\eta }\right) ^{\gamma }\right) \left( \frac{\mathrm{e}^{y}}{%
\eta }\right) ^{\gamma }
\end{equation*}%
where, as before, $0<\gamma <1$ and $\eta >0$. With the previous change of
variable, the lognormal becomes a normal distribution (N), the mixtures 2LN
and 3LN become, in an obvious notation, the mixtures 2N and 3N, the power
law or Pareto distribution becomes an exponential distribution like
\begin{equation*}
f_{\mathrm{EXP}}(y;\alpha ,y_{\mathrm{min}})=(\alpha -1)\exp \left( -(\alpha
-1)(y-y_{\mathrm{min}})\right)
\end{equation*}%
where $y_{\mathrm{min}}=\ln (x_{\mathrm{min}})$ and $y_{\mathrm{min}}\leq y$%
. The truncated lognormal becomes, likewise, a truncated normal:
\begin{equation*}
f_{\mathrm{Nt}}(y;\mu ,\sigma ,y_{\mathrm{min}})=\frac{f_{\mathrm{N}}(y;\mu
,\sigma )}{1-\mathrm{cdf}_{\mathrm{N}}(y_{\mathrm{min}};\mu ,\sigma )}
\end{equation*}%
where, obviously, $y_{\mathrm{min}}$ is as in the previous exponential
distribution with $y_{\mathrm{min}}\leq y$, and
\begin{eqnarray}
&&f_{\mathrm{N}}(y;\mu ,\sigma )=\frac{1}{\sqrt{2\pi }\sigma }\exp \left( -%
\frac{(y-\mu )^{2}}{2\sigma ^{2}}\right)  \notag \\
&&\mathrm{cdf}_{\mathrm{N}}(y;\mu ,\sigma )=\frac{1}{2}+\frac{1}{2}\mathrm{%
erf}\left( \frac{y-\mu }{\sqrt{2}\sigma }\right)  \notag
\end{eqnarray}%
being now $\mu \in (-\infty ,\infty )$.

We consider stochastic models whose stationary density functions are the
distributions just described. Let $y_{t}$ denote a random variable, which in
principle can depend on time, whose evolution or dynamics is governed by the
It\^{o} differential equation (see, e.g., \cite{Ord74,Gar04})
\begin{equation}
dy_{t}=b(y_{t},t)dt+\sqrt{a(y_{t},t)}dB_{t}  \label{ito}
\end{equation}%
where $B_{t}$ is a standard Brownian motion (Wiener process) (see, e.g., %
\cite{ItoMcK96,Kyp06} and references therein). The quantity $a(y_{t},t)$
corresponds to the \emph{diffusion term}, and $b(y_{t},t)$ to the \emph{%
drift term}. This process can be associated to the \emph{forward Kolmogorov
equation} or \emph{Fokker-Planck equation} for the time-dependent
probability density function (conditional on the initial data) $f(y,t)$ (see
also \cite{Gab99,Gab09}):
\begin{equation*}
\frac{\partial f(y,t)}{\partial t}=-\frac{\partial }{\partial y}%
(b(y,t)f(y,t))+\frac{1}{2}\frac{\partial ^{2}}{\partial y^{2}}(a(y,t)f(y,t)).
\end{equation*}

Like most of the literature (see, e.g., \cite%
{FatLem06,GuaTos18,GuaTos19,GuaTos19b,MogMilSer20}), we work with stationary probability
density functions, so that for different choices of (time-independent) $b(y)$
and $a(y)$, the Fokker--Planck equation can be solved and produce the
distributions we consider \cite%
{Ord74,Dup93,FatLem06,FurPulTerTos17,DolLeoOut17,GuaTos19} For example, the
``exponential stretched exponential''\ is a stationary solution of the
Fokker--Planck equation with the choice $a(y)=\sigma ^{2}$, where $\sigma >0$
is a constant, and
\begin{equation*}
b(y)=-\frac{\gamma \sigma ^{2}}{2}\left( \left( \frac{\mathrm{e}^{y}}{\eta }%
\right) ^{\gamma }-1\right)
\end{equation*}%
Stretched exponential distributions can be generated as well by
multiplicative processes with limiting behavior determined by the
application of extreme deviations theory \citep{LahSor98,FriSor97}. The
recent work \cite{Cam20} discussed how extreme deviations theory applied to
a proportionate growth model, which is the deterministic version of the
Brownian motion in equation (\ref{ito}), will generate a stretched
exponential distribution. Likewise, the normal distribution can be obtained
as a stationary solution to Fokker--Planck equation by assuming the
diffusion term is given as well by a constant \citep{UhlOrn30,Kal45,Vas77} $%
a(y)=\sigma ^{2}$ and $b(y)=-\frac{1}{2}(y-\mu)$, exhibiting \emph{mean
reversion}. The exponential distribution can also be obtained as a
stationary solution to the Fokker--Planck equation with drift and diffusion
terms given by $a(y)=\sigma ^{2}$, ${\displaystyle b(y)=-\frac{1}{2}(\alpha
-1)\sigma ^{2}}$. We can also obtain the truncated normal distribution as a
stationary solution to the Fokker--Planck equation, with drift and diffusion
terms $a(y)=\sigma ^{2}$, ${\displaystyle b(y)=-\frac{1}{2}(y-\mu )}$. This
is similar to our solution for the normal equation distribution pointed out
earlier, except that now the support of the distribution is $[y_{\mathrm{min}%
},\infty )$.

We can also obtain mixtures of normal distributions, our (exponentiated)
models $f_{\mathrm{2N}}$, $f_{\mathrm{3N}}$ (mixtures of two or three
normals, respectively) as stationary solutions to the Fokker--Planck
equations and formulate stochastic equations for the strike processes. We
first consider $f_{\mathrm{2N}}$, a mixture of two normal distributions,
given by
\begin{equation*}
f_{\mathrm{2N}}(y;\mu _{1},\sigma _{1},\mu _{2},\sigma _{2},p_{1})=p_{1}f_{%
\mathrm{N}}(y;\mu _{1},\sigma _{1})+(1-p_{1})f_{\mathrm{N}}(y;\mu
_{2},\sigma _{2})
\end{equation*}%
where $0\leq p_{1},1-p_{1}\leq 1$. We will denote for simplicity the \emph{%
posterior probabilities} (see, e.g., \cite{McLPee00})
\begin{eqnarray}
&&\tau _{1}(y)=p_{1}f_{\mathrm{N}}(y;\mu _{1},\sigma _{1})/f_{\mathrm{2N}%
}(y;\mu _{1},\sigma _{1},\mu _{2},\sigma _{2},p_{1})  \notag \\
&&\tau _{2}(y)=(1-p_{1})f_{\mathrm{N}}(y;\mu _{2},\sigma _{2})/f_{\mathrm{2N}%
}(y;\mu _{1},\sigma _{1},\mu _{2},\sigma _{2},p_{1})  \notag
\end{eqnarray}%
Then, let the diffusion and drift terms be defined by $a(y)=\sigma^{2}$, and
\begin{equation*}
b(y)=-\frac{\sigma ^{2}}{2\sigma _{1}^{2}}\tau _{1}(y)(y-\mu _{1})-\frac{%
\sigma ^{2}}{2\sigma _{2}^{2}}\tau _{2}(y)(y-\mu _{2})
\end{equation*}%
so that we obtain that the $f_{\mathrm{2N}}$ above is a stationary solution
of the corresponding Fokker--Planck equation. This is a novel, as far as we
know, generalization of the results of \cite{UhlOrn30,Kal45,Vas77}, where
the relevant quantities can be functions of $y$ with definite sign. For the $%
f_{\mathrm{3N}}$, that is, a mixture of three normal distributions, it is
completely analogous to the case of 2N. The 3N can be given by
\begin{equation*}
f_{\mathrm{3N}}(y;\mu _{1},\sigma _{1},\mu _{2},\sigma _{2},\mu _{3},\sigma
_{3},p_{1},p_{2})=p_{1}f_{\mathrm{N}}(y;\mu _{1},\sigma _{1})+p_{2}f_{%
\mathrm{N}}(y;\mu _{2},\sigma _{2})+(1-p_{1}-p_{2})f_{\mathrm{N}}(y;\mu
_{3},\sigma _{3})
\end{equation*}%
where $0\leq p_{1},p_{2},1-p_{1}-p_{2}\leq 1$. We define in this case the
corresponding posterior probabilities
\begin{eqnarray}
&&\tau _{1}(y)=p_{1}f_{\mathrm{N}}(y;\mu _{1},\sigma _{1})/f_{\mathrm{3N}%
}(y;\mu _{1},\sigma _{1},\mu _{2},\sigma _{2},\mu _{3},\sigma
_{3},p_{1},p_{2})  \notag \\
&&\tau _{2}(y)=p_{2}f_{\mathrm{N}}(y;\mu _{2},\sigma _{2})/f_{\mathrm{3N}%
}(y;\mu _{1},\sigma _{1},\mu _{2},\sigma _{2},\mu _{3},\sigma
_{3},p_{1},p_{2})  \notag \\
&&\tau _{3}(y)=(1-p_{1}-p_{2})f_{\mathrm{N}}(y;\mu _{3},\sigma _{3})/f_{%
\mathrm{3N}}(y;\mu _{1},\sigma _{1},\mu _{2},\sigma _{2},\mu _{3},\sigma
_{3},p_{1},p_{2})  \notag
\end{eqnarray}%
and the diffusion and drift terms defined by $a(y)=\sigma ^{2}$,
\begin{equation*}
b(y)=-\frac{\sigma ^{2}}{2\sigma _{1}^{2}}\tau _{1}(y)(y-\mu _{1})-\frac{%
\sigma ^{2}}{2\sigma _{2}^{2}}\tau _{2}(y)(y-\mu _{2})-\frac{\sigma ^{2}}{%
2\sigma _{3}^{2}}\tau _{3}(y)(y-\mu _{3})
\end{equation*}%
we obtain that the $f_{\mathrm{3N}}$ above is a stationary solution of the
corresponding Fokker--Planck equation.

The Fokker--Planck equations are the continuous time counterparts to the
dynamics that arise in kinetic models. These kinetic models have been
increasingly used to study socioeconomic and human behavior %
\citep{GuaTos18,GuaTos19,GuaTos19b,DolLeoOut17,FurPulTerTos17}. These models
may also be applicable when considering processes generating strikes, which
arise during collective bargaining, i.e., the interactions between the
unions, the bargaining agent for workers, with management. In particular,
there are individual agents, which are comprised of firm and union
representatives who form a bargaining pair. The collective bargaining that
occurs between these bargaining pairs is analogous to the interactions
between particles of gases. The system is compromised of negotiations by
other bargaining pairs and they can view average settlement terms as well as
specific contracts. While contract negotiations often end in a settlement
they can sometimes result in an impasse and, consequently, a strike occurs
which can vary in size (person days lost). The stationary solutions to the
Fokker--Planck equations we described can provide alternative distributions
for the size of the strikes that occur in this kinetic formulation of
collective bargaining. For example, we can obtain exponential or normal
distributions for the log-sizes (power law or lognormal distributions for
the sizes, respectively) with suitable choices of drift and diffusion terms,
which reflect different conditions on the interactions between bargaining
pairs. Moreover, if we were to split the agents (bargaining pairs) into
different groups by the log-size of the bargaining pair (log of the number
of persons involved), which creates subsystems of agents, then we can obtain
mixtures of normal distributions (lognormal for the sizes). This result was
also derived in \cite{GuaTos19} for the distribution of city sizes. The
kinetic model can thus lay the framework that creates the Fokker--Planck
equations whose stationary solutions are the distribution of strike size.

\section{The databases}

\label{databases}

The data used in this analysis are compiled from a few different sources and
cover two periods: the years prior to World War II; and, the post-World War
II period up until the mid-1970s. While aggregated strike data is available
in many countries, micro data on strikes is less common. Since our objective
is to study the distribution of strike size we require micro data on
individual strikes. We should also note that our strike data also includes
lockouts, where the work stoppage was initiated by the firm, so when we use
the term strike it would include both types of work stoppages (i.e., worker
and firm initiated).

For the pre-World War II period, we have data for France (all industries,
1890-1935), Canada (all industries,1901-1916), the Netherlands (all
industries, 1890-1935) and the United States (1881-1886 and 1887-1894) all
industries for 3 selected states. The U.S. data we compiled includes strikes
and lockouts for 3 selected states, one from each of the major economic
regions in the U.S., the mid-west (Illinois), New England (Massachusetts)
and the mid-Atlantic (New Jersey). Information on these strikes was
collected in the Commissioner of Labor's 3rd and 10th Annual reports %
\citep{us1888,us1896}. We keep the data for the 1881-1886 and 1887-1894
separated because there were some changes in data collection procedures in
terms of how strikes were defined between the 3rd annual report (1881-1886)
and the 10th annual report (1887-1894) so the data may not be comparable. In
particular, the 3rd annual report used the firm as the unit of observation
so that related strikes at different plants (workplaces) in the firm would
be counted as separate strikes. The 10th annual report used a broader
definition of a strike, which would count strikes over separate issues at
different plants as a single dispute instead of reporting them as individual
strikes. For the post-World War II period, we have data on Canada
(1946-1975), the Netherlands (1946-1975) and the United States (1953-1977).
The details on the sources of all these data are provided in the Data
Appendix.

We measure strike size as person days lost, which is computed as the number
of persons on strike multiplied by strike duration. This measure is a two
dimensional measure of strike size that addresses the limitations of using
its components individually. For example, the number of persons on strike
does not capture any variation in strike duration and, similarly, strike
duration does not provide any information on how many participants were
involved. For example, \cite{Cam19b} discussed that in his data the 10
longest strikes lasted at least 1500 days and 3 of the 10 longest strikes
involved less than 20 workers. Thus the person days lost captures both types
of information, which would be missed in the one-dimensional measures that
are used to create it.

We consider two samples: the full sample, which includes, all the data we
have collected; and, truncated samples only considering the upper tails of
the distributions.

The minimum cut-off values $x_{\mathrm{min}}$ for the truncated samples are
obtained as follows. We rely on an approach due to \cite{ClaYouSke07}. Their
approach is based on the idea of picking $x_{\mathrm{min}}$ to make the
difference between the observed data and the fitted power law distribution
as small as possible. This can be operationalized by using the
Kolmogorov--Smirnov (KS) statistic,
\begin{equation}
D=\underset{x\geq x_{\mathrm{min}}}{\max }\left| S(x)-P(x)\right|
\label{ksstat}
\end{equation}%
where $S(x)$ is the empirical cumulative distribution function (CDF) of the
data for observations with a value of at least $x_{\mathrm{min}}$ and $P(x)$
is the CDF of the power law distribution that best fits the data with $x\geq
x_{\mathrm{min}}$. Thus our estimate of $x_{\mathrm{min}}$ minimizes $D$ in
equation (\ref{ksstat}).

We present descriptive statistics in Table~\ref{descstat} for strike size
for each country and period we consider.

\begin{table}[htbp]
\centering {\tiny
\begin{tabular}{llllllllll}
\toprule Lost person days & Sample size & Mean & SD & Mean (log scale) & SD
(log scale) & Skewness (log scale) & Kurtosis (log scale) & Min & Max \\
Canada 1901-1916 & 1238 & 7665 & 51496 & 6.548 & 2.007 & 0.282 & 3.079 & 6 &
1400000 \\
Canada 1945-1975 & 11465 & 8767 & 54391 & 6.631 & 2.125 & 0.25 & 2.798 & 10
& 1600000 \\
Netherlands 1890-1935 & 7318 & 3717 & 45428 & 5.161 & 2.256 & 0.299 & 3.071
& 1 & 2700000 \\
Netherlands 1946-1975 & 2007 & 2685 & 26820 & 4.71 & 2.026 & 0.694 & 3.918 &
1 & 666000 \\
France 1890-1935 & 16247 & 5980 & 89649 & 6.063 & 1.954 & 0.422 & 3.322 & 1
& 8001168 \\
USA 1881-1886 & 1301 & 7699 & 35928 & 6.699 & 2.106 & 0.189 & 2.784 & 3 &
846055 \\
USA 1886-1894 & 2309 & 7320 & 65899 & 5.789 & 2.187 & 0.551 & 3.279 & 1 &
1915420 \\
USA 1953-1977 & 63562 & 6175 & 96305 & 6.405 & 1.974 & 0.283 & 3.006 & 1 &
17097371 \\
&  &  &  &  &  &  &  &  &  \\
Lost person days, upper tail & Sample size & Mean & SD & Mean (log scale) &
SD (log scale) & Skewness (log scale) & Kurtosis (log scale) & Min & Max \\
Canada 1901-1916 & 247 & 35313 & 111232 & 9.424 & 1.109 & 1.63 & 5.519 & 4000
& 1400000 \\
Canada 1945-1975 & 566 & 128524 & 210941 & 11.227 & 0.871 & 1.341 & 4.414 &
29000 & 1600000 \\
Netherlands 1890-1935 & 1145 & 22325 & 113082 & 8.78 & 1.18 & 1.412 & 5.173
& 1800 & 2700000 \\
Netherlands 1946-1975 & 423 & 12361 & 57449 & 7.671 & 1.404 & 1.48 & 5.168 &
500 & 666000 \\
France 1890-1935 & 374 & 188400 & 561512 & 11.356 & 0.958 & 1.636 & 5.97 &
31680 & 8001168 \\
USA 1881-1886 & 197 & 44156 & 83507 & 10.097 & 0.936 & 1.099 & 3.905 & 8050
& 846055 \\
USA 1886-1894 & 639 & 25857 & 123426 & 8.568 & 1.401 & 1.226 & 4.342 & 1050
& 1915420 \\
USA 1953-1977 & 1157 & 198136 & 685888 & 11.655 & 0.765 & 1.766 & 7.78 &
51000 & 17097371 \\
\bottomrule &  &  &  &  &  &  &  &  &
\end{tabular}%
}
\caption{Descriptive statistics of the full and upper tail samples}
\label{descstat}
\end{table}

\section{Results}

\label{results}

We report the ML parameter estimates for the stretched exponential (STEXP),
2LN, 3LN (full samples) and the Pareto, LNt (upper tail samples),
respectively, and the corresponding standard errors (SE) in Tables~\ref%
{estim2ln3ln} and~\ref{estimplnt}.

\begin{table}[htbp]
\centering {\tiny
\begin{tabular}{lllllllll}
\toprule & STEXP &  &  &  &  &  &  &  \\
& $\gamma$ (SE) & $\eta$ (SE) &  &  &  &  &  &  \\
Canada 1901-1916 & 0.465 (0.009) & 1947.839 (119.078) &  &  &  &  &  &  \\
Canada 1945-1975 & 0.452 (0.003) & 2242.004 (46.354) &  &  &  &  &  &  \\
Netherlands 1890-1935 & 0.414 (0.003) & 553.773 (15.639) &  &  &  &  &  &
\\
Netherlands 1946-1975 & 0.417 (0.006) & 324.131 (17.347) &  &  &  &  &  &
\\
France 1890-1935 & 0.459 (0.002) & 1178.591 (20.139) &  &  &  &  &  &  \\
USA 1881-1886 & 0.465 (0.009) & 2365.235 (141.154) &  &  &  &  &  &  \\
USA 1886-1894 & 0.408 (0.005) & 1025.185 (52.251) &  &  &  &  &  &  \\
USA 1953-1977 & 0.476 (0.001) & 1659.852 (13.833) &  &  &  &  &  &  \\
&  &  &  &  &  &  &  &  \\
& 2LN &  &  &  &  &  &  &  \\
& $\mu_1$ (SE) & $\sigma_1$ (SE) & $\mu_2$ (SE) & $\sigma_2$ (SE) & $p_1$
(SE) &  &  &  \\
Canada 1901-1916 & 8.601 (0.290) & 2.178 (0.201) & 6.291 (0.060) & 1.827
(0.044) & 0.111 (0.020) &  &  &  \\
Canada 1945-1975 & 7.363 (0.026) & 1.972 (0.018) & 4.954 (0.035) & 1.381
(0.024) & 0.696 (0.008) &  &  &  \\
Netherlands 1890-1935 & 6.397 (0.054) & 2.239 (0.037) & 4.269 (0.036) &
1.803 (0.026) & 0.419 (0.012) &  &  &  \\
Netherlands 1946-1975 & 6.662 (0.146) & 2.277 (0.101) & 4.141 (0.045) &
1.534 (0.034) & 0.226 (0.016) &  &  &  \\
France 1890-1935 & 7.246 (0.034) & 2.019 (0.024) & 5.344 (0.019) & 1.514
(0.014) & 0.378 (0.008) &  &  &  \\
USA 1881-1886 & 7.468 (0.082) & 1.970 (0.058) & 5.285 (0.105) & 1.536 (0.073)
& 0.648 (0.026) &  &  &  \\
USA 1886-1894 & 7.280 (0.094) & 2.195 (0.065) & 4.798 (0.051) & 1.516 (0.038)
& 0.399 (0.017) &  &  &  \\
USA 1953-1977 & 7.396 (0.014) & 1.925 (0.010) & 5.511 (0.011) & 1.543 (0.008)
& 0.474 (0.004) &  &  &  \\
&  &  &  &  &  &  &  &  \\
& 3LN &  &  &  &  &  &  &  \\
& $\mu_1$ (SE) & $\sigma_1$ (SE) & $\mu_2$ (SE) & $\sigma_2$ (SE) & $\mu_3$
(SE) & $\sigma_3$ (SE) & $p_1$ (SE) & $p_2$ (SE) \\
Canada 1901-1916 & 4.611 (0.089) & 1.194 (0.063) & 12.184 (0.227) & 0.796
(0.167) & 7.178 (0.066) & 1.599 (0.052) & 0.276 (0.019) & 0.016 (0.004) \\
Canada 1945-1975 & 3.372 (0.038) & 0.738 (0.025) & 5.647 (0.034) & 1.290
(0.026) & 7.756 (0.030) & 1.890 (0.021) & 0.078 (0.003) & 0.371 (0.008) \\
Netherlands 1890-1935 & 6.782 (0.063) & 2.201 (0.044) & 4.764 (0.038) &
1.707 (0.029) & 2.409 (0.079) & 1.248 (0.053) & 0.317 (0.006) & 0.580 (0.007)
\\
Netherlands 1946-1975 & 7.196 (0.182) & 2.285 (0.129) & 5.926 (0.113) &
1.166 (0.098) & 3.720 (0.045) & 1.334 (0.033) & 0.153 (0.012) & 0.207 (0.016)
\\
France 1890-1935 & 8.510 (0.095) & 2.174 (0.065) & 6.640 (0.026) & 1.662
(0.020) & 4.804 (0.023) & 1.335 (0.017) & 0.081 (0.004) & 0.523 (0.007) \\
USA 1881-1886 & 8.734 (0.104) & 1.609 (0.074) & 7.115 (0.100) & 0.874 (0.085)
& 5.038 (0.071) & 1.362 (0.051) & 0.318 (0.015) & 0.233 (0.018) \\
USA 1886-1894 & 8.420 (0.140) & 2.068 (0.100) & 5.934 (0.071) & 1.566 (0.057)
& 3.962 (0.065) & 1.214 (0.047) & 0.183 (0.010) & 0.513 (0.017) \\
USA 1953-1977 & 7.641 (0.016) & 1.893 (0.011) & 5.872 (0.012) & 1.433 (0.009)
& 3.630 (0.023) & 0.906 (0.016) & 0.396 (0.002) & 0.530 (0.002) \\
\bottomrule &  &  &  &  &  &  &  &
\end{tabular}%
}
\caption{ML estimators and Standard Errors (SE) for the whole samples and
the STEXP, 2LN and 3LN distributions. The estimators of the LN are the
sample mean and standard deviations of the log-data, see Table~\ref{descstat}%
.}
\label{estim2ln3ln}
\end{table}

\begin{table}[htbp]
\centering {\tiny
\begin{tabular}{llll}
\toprule & Pareto & LNt &  \\
& $\alpha$ (SE) & $\mu$ (SE) & $\sigma$ (SE) \\
Canada 1901-1916 & 1.885 (0.056) & -40.997 (3.350) & 7.632 (0.248) \\
Canada 1945-1975 & 2.051 (0.044) & 4.239 (0.358) & 2.721 (0.063) \\
Netherlands 1890-1935 & 1.779 (0.023) & -1.286 (0.359) & 3.784 (0.061) \\
Netherlands 1946-1975 & 1.687 (0.033) & -24.619 (1.698) & 7.000 (0.177) \\
France 1890-1935 & 2.007 (0.052) & -11.049 (1.252) & 4.813 (0.129) \\
USA 1881-1886 & 1.906 (0.065) & 7.027 (0.325) & 2.064 (0.090) \\
USA 1886-1894 & 1.620 (0.025) & 2.596 (0.327) & 3.404 (0.079) \\
USA 1953-1977 & 2.227 (0.036) & 2.337 (0.315) & 2.860 (0.045) \\
\bottomrule &  &  &
\end{tabular}%
}
\caption{ML estimators and Standard Errors (SE) for the truncated samples
(upper tails) and the Pareto, LNt distributions.}
\label{estimplnt}
\end{table}

The ML estimators for the parameters of the log normal distribution are
presented in Table~\ref{descstat}. We observe that in all of the shown
estimations, the parameters are statistically significant at the 5\% level.

\begin{figure}[tbp]
\centering
\subfloat[][]{\includegraphics[width=.5\textwidth,
height=5cm]{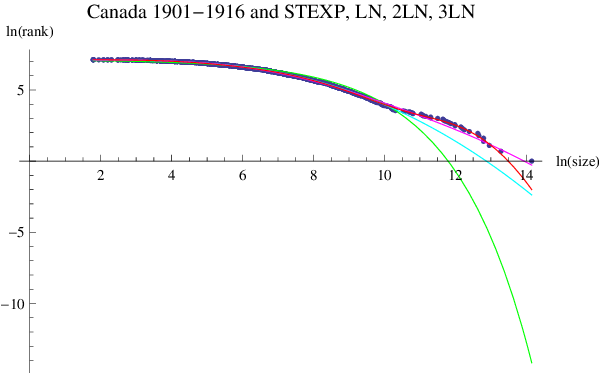}} \hfill
\subfloat[][]{\includegraphics[width=.5\textwidth,
  height=5cm]{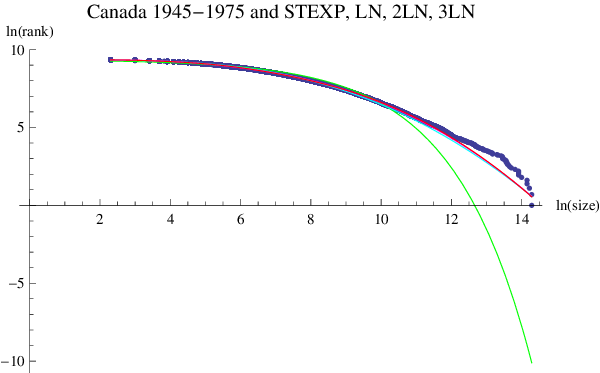}}
\par
\subfloat[][]{\includegraphics[width=.45\textwidth,
height=5cm]{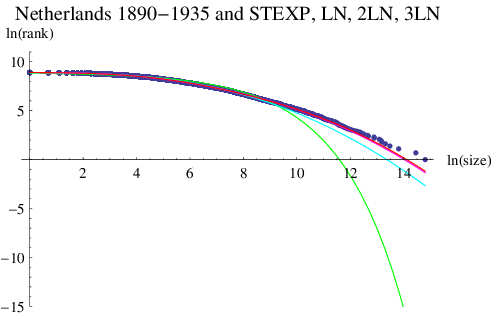}} \hfill
\subfloat[][]{\includegraphics[width=.45\textwidth,
  height=5cm]{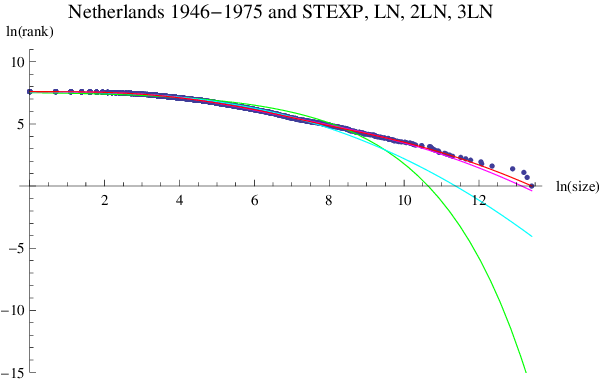}}
\par
\subfloat[][]{\includegraphics[width=.5\textwidth, height=5cm]{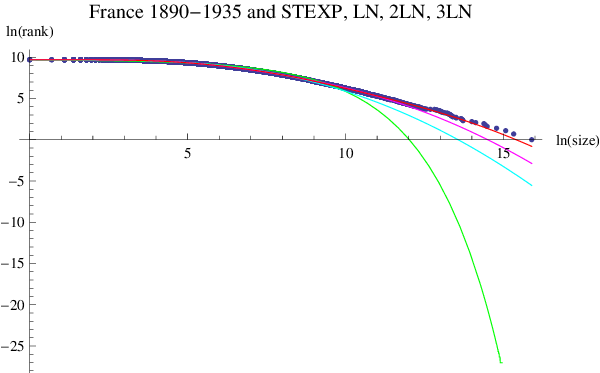}}
\hfill
\subfloat[][]{\includegraphics[width=.5\textwidth,
  height=5cm]{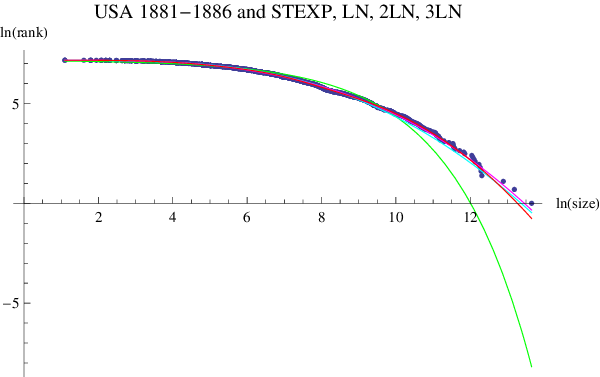}}
\par
\subfloat[][]{\includegraphics[width=.5\textwidth, height=5cm]{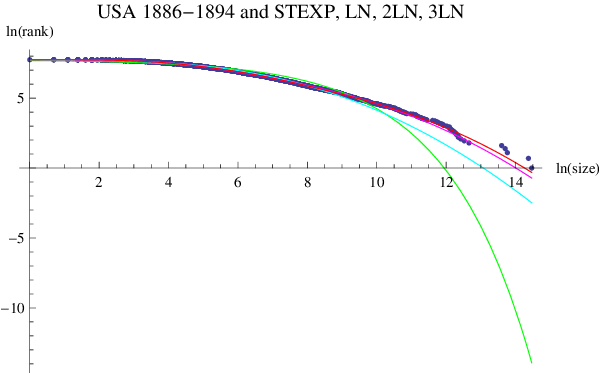}}
\hfill
\subfloat[][]{\includegraphics[width=.5\textwidth,
  height=5cm]{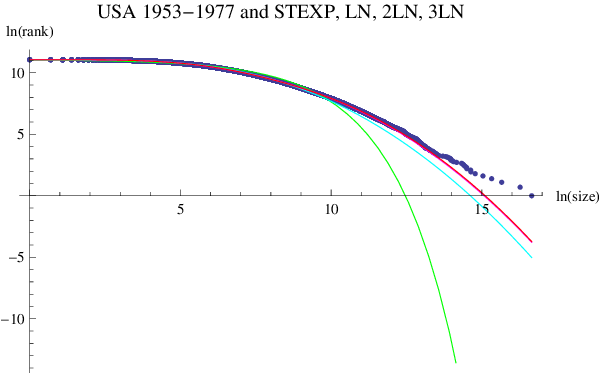}}
\caption{Log-rank plots for the whole samples of log-sizes, using the STEXP
(green), LN (cyan), 2LN (magenta) and 3LN (red) distributions and the
empirical data (blue). (In color online).}
\label{r3l}
\end{figure}

\begin{figure}[tbp]
\centering
\subfloat[][]{\includegraphics[width=.5\textwidth,
height=5cm]{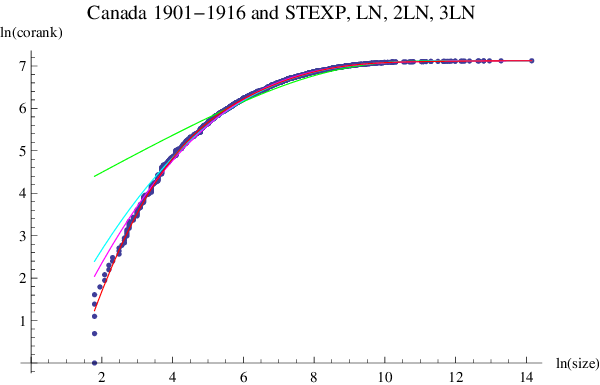}} \hfill
\subfloat[][]{\includegraphics[width=.5\textwidth,
  height=5cm]{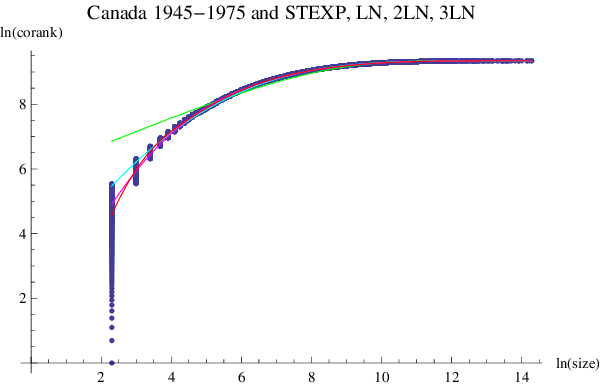}}
\par
\subfloat[][]{\includegraphics[width=.5\textwidth,
height=5cm]{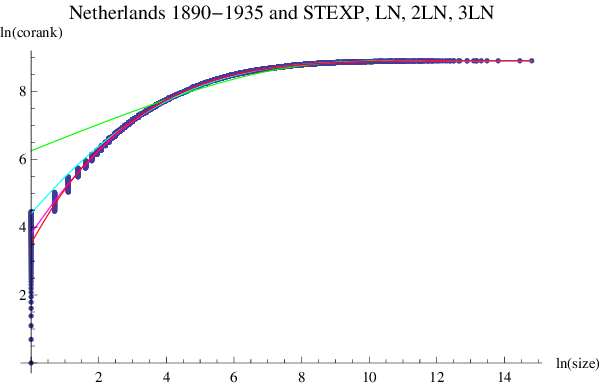}} \hfill
\subfloat[][]{\includegraphics[width=.5\textwidth,
  height=5cm]{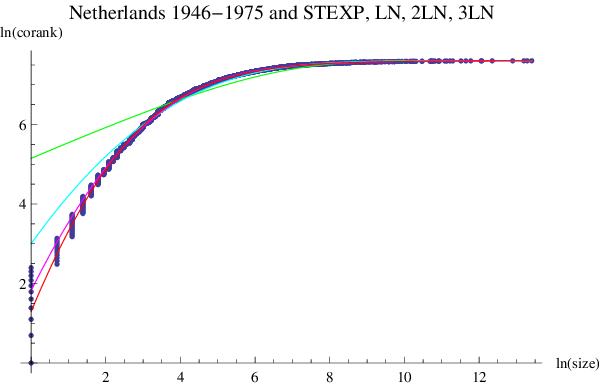}}
\par
\subfloat[][]{\includegraphics[width=.5\textwidth, height=5cm]{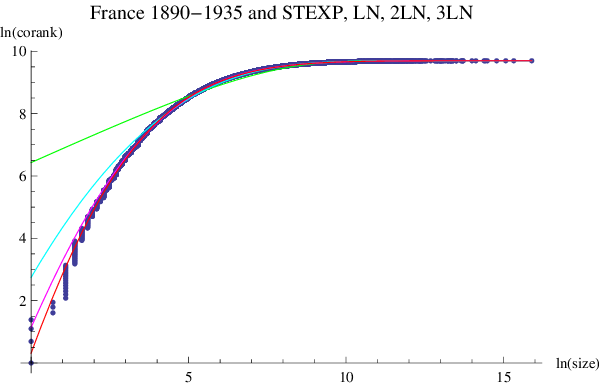}}
\hfill
\subfloat[][]{\includegraphics[width=.5\textwidth,
  height=5cm]{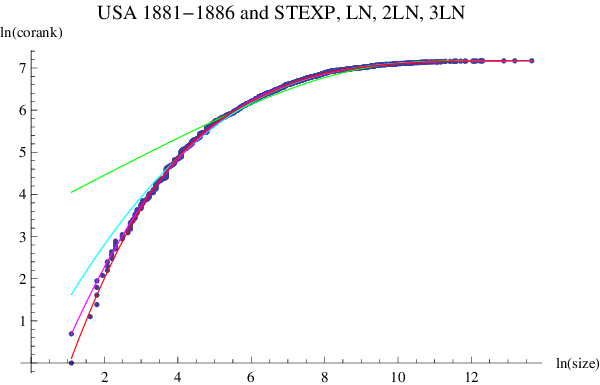}}
\par
\subfloat[][]{\includegraphics[width=.5\textwidth,
height=5cm]{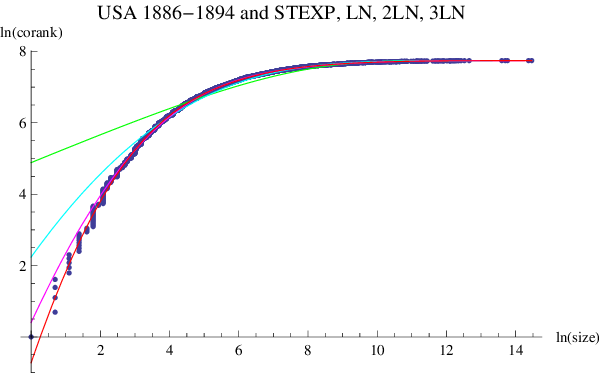}} \hfill
\subfloat[][]{\includegraphics[width=.5\textwidth,
  height=5cm]{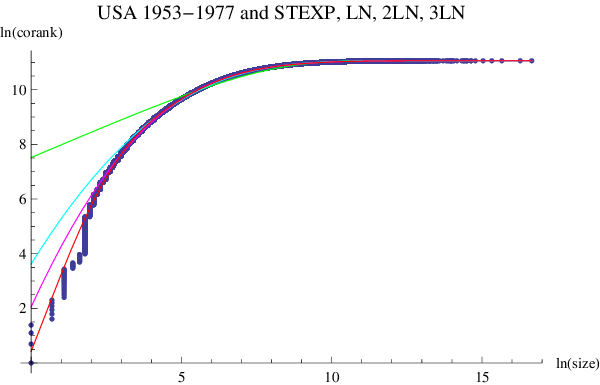}}
\caption{Log-corank plots for the whole samples of log-sizes, using the
STEXP (green), LN (cyan), 2LN (magenta) and 3LN (red) distributions and the
empirical data (blue). (In color online).}
\label{cr3l}
\end{figure}

\begin{figure}[tbp]
\centering
\subfloat[][]{\includegraphics[width=.5\textwidth,
height=5cm]{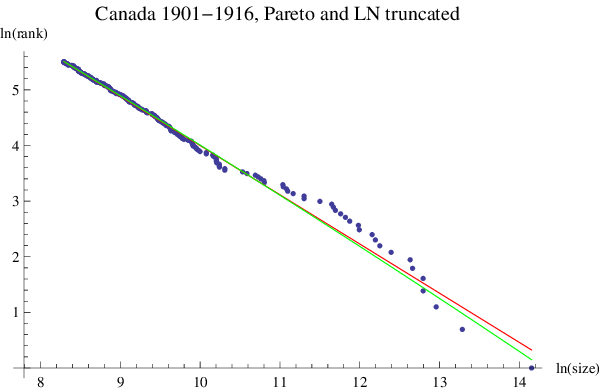}} \hfill
\subfloat[][]{\includegraphics[width=.5\textwidth,
  height=5cm]{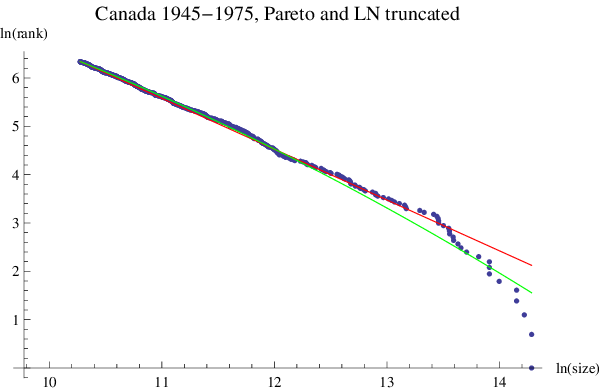}}
\par
\subfloat[][]{\includegraphics[width=.5\textwidth,
height=5cm]{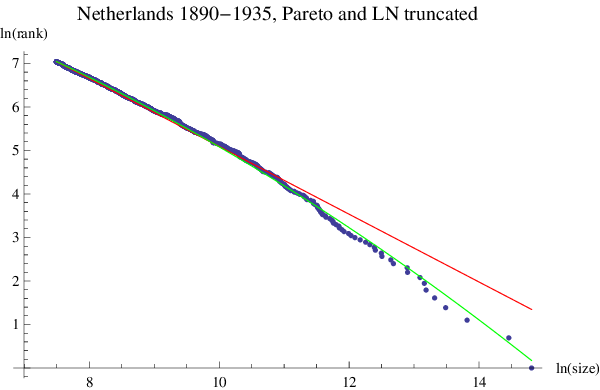}} \hfill
\subfloat[][]{\includegraphics[width=.5\textwidth,
  height=5cm]{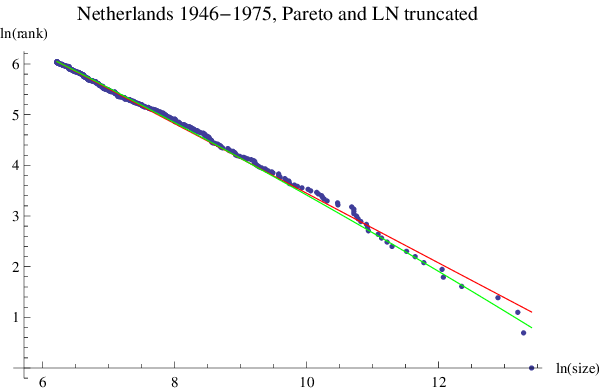}}
\par
\subfloat[][]{\includegraphics[width=.5\textwidth, height=5cm]{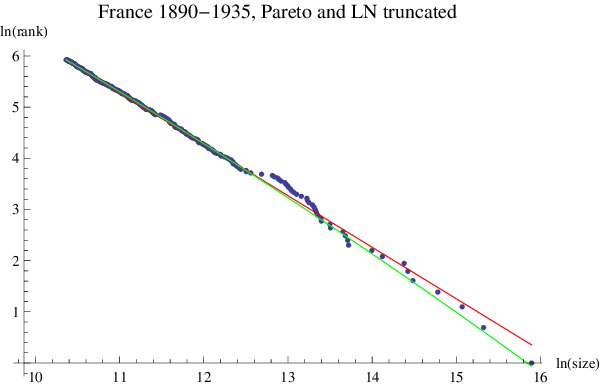}}
\hfill
\subfloat[][]{\includegraphics[width=.5\textwidth,
  height=5cm]{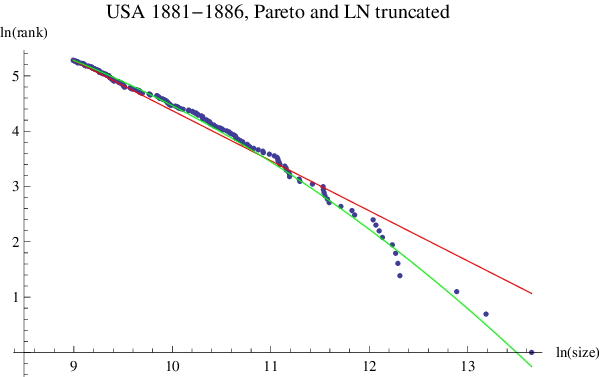}}
\par
\subfloat[][]{\includegraphics[width=.5\textwidth,
height=5cm]{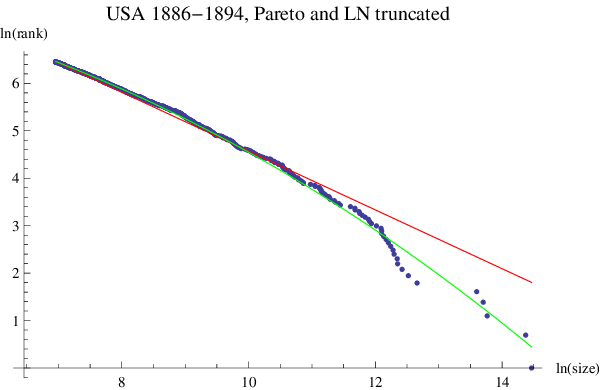}} \hfill
\subfloat[][]{\includegraphics[width=.5\textwidth,
  height=5cm]{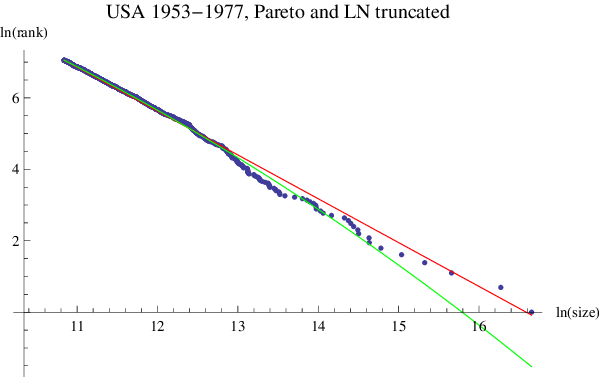}}
\caption{Log-rank plots for the upper tails, using the power law or Pareto
distribution (red), the truncated lognormal LNt (green) and the empirical
data (blue). (In color online).}
\label{rput}
\end{figure}

We first consider a graphical approach to assessing goodness-of-fit with
log-rank/corank plots. We have plotted these quantities for all the models
we consider. The deviations at the upper (resp. lower) tails are amplified
since we use logarithms (see, e.g., \cite{GonRamSan13}). As is well known,
the plot of the upper tails should be linear if the distribution in the
tails is Pareto. We present these plots in Figures~\ref{r3l}, \ref{cr3l} and~%
\ref{rput}, i.e., the plots of log-ranks, log-coranks for the whole samples
and log-ranks for the upper tail samples, respectively.

In Figures~\ref{r3l} and~\ref{cr3l}, we observe that the fit of the STEXP is
rather poor, meanwhile those of the 2LN and 3LN are really good (small
discrepancies), and the LN performs a little worse than the second ones
(larger discrepancies). We also observe a great deal of curvature in the
data. In Figure~\ref{rput}, the discrepancies are also small and the data
are approximately linear, making very hard to assess whether the Pareto or
the truncated lognormal provides the best fit. Figure~\ref{rput} illustrates
the difficulty of distinguishing Pareto or power laws for the upper tails
{}from other alternatives %
\citep{Per05,BeeRicSch11,BeeRicSch13,BeeRicSch19,SchTre18}.

\begin{table}[htbp]
\centering {\tiny
\begin{tabular}{lllllllrrr}
\toprule & STEXP &  &  &  &  &  &  &  &  \\
& KS & CM & AD &  &  &  &  &  &  \\
Canada 1901-1916 & 0 (0..088) & 0 (2.987) & 0 (21.965) &  &  &  &  &  &  \\
Canada 1945-1975 & 0 (0.089) & 0 (23.456) & 0 (167.264) &  &  &  &  &  &  \\
Netherlands 1890-1935 & 0 (0.090) & 0 (19.660) & 0 (134.767) &  &  &  &  &
&  \\
Netherlands 1946-1975 & 0 (0.129) & 0 (11.597) & 0 (74.266) &  &  &  &  &  &
\\
France 1890-1935 & 0 (0.103) & 0 (51.602) & 0 (358.700) &  &  &  &  &  &  \\
USA 1881-1886 & 0 (0.082) & 0 (3.047) & 0 (20.656) &  &  &  &  &  &  \\
USA 1886-1894 & 0 (0.110) & 0 (10.157) & 0 (65.858) &  &  &  &  &  &  \\
USA 1953-1977 & 0 (0.084) & 0 (109.540) & 0 (824.416) &  &  &  &  &  &  \\
&  &  &  &  &  &  &  &  &  \\
& LN &  &  & 2LN &  &  & \multicolumn{1}{l}{3LN} &  &  \\
& KS & CM & AD & KS & CM & AD & \multicolumn{1}{l}{KS} & \multicolumn{1}{l}{
CM} & \multicolumn{1}{l}{AD} \\
Canada 1901-1916 & \textbf{0.513 (0.023)} & \textbf{0.526 (0.113)} & \textbf{%
0.396 (0.931)} & \textbf{0.582 (0.022)} & \textbf{0.526 (0.113)} & \textbf{%
0.492 (0.785)} & \multicolumn{1}{l}{\textbf{0.893 (0.016)}} &
\multicolumn{1}{l}{\textbf{0.990 (0.025)}} & \multicolumn{1}{l}{\textbf{%
0.995 (0.181)}} \\
Canada 1945-1975 & 0 (0.029) & 0 (2.172) & 0 (13.640) & 0.019 (0.015) &
\textbf{0.192 (0.247)} & 0.031 (2.885) & \multicolumn{1}{l}{0.028 (0.014)} &
\multicolumn{1}{l}{\textbf{0.464 (0.128)}} & \multicolumn{1}{l}{\textbf{%
0.059 (2.359)}} \\
Netherlands 1890-1935 & 0.001 (0.024) & 0.004 (0.923) & 0.001 (6.333) &
\textbf{0.251 (0.012)} & \textbf{0.572 (0.103)} & \textbf{0.386 (0.947)} &
\multicolumn{1}{l}{\textbf{0.579 (0.009)}} & \multicolumn{1}{l}{\textbf{%
0.664 (0.085)}} & \multicolumn{1}{l}{\textbf{0.475 (0.809)}} \\
Netherlands 1946-1975 & 0 (0.056) & 0 (1.675) & 0 (10.548) & 0.030 (0.033) &
\textbf{0.393 (0.149)} & \textbf{0.462 (0.827)} & \multicolumn{1}{l}{\textbf{%
0.402 (0.020)}} & \multicolumn{1}{l}{\textbf{0.899 (0.046)}} &
\multicolumn{1}{l}{\textbf{0.932 (0.308)}} \\
France 1890-1935 & 0 (0.035) & 0 (4.560) & 0 (27.218) & \textbf{0.082 (0.011)%
} & \textbf{0.359 (0.161)} & \textbf{0.258 (1.225)} & \multicolumn{1}{l}{%
\textbf{0.302 (0.008)}} & \multicolumn{1}{l}{\textbf{0.814 (0.060)}} &
\multicolumn{1}{l}{\textbf{0.861 (0.388)}} \\
USA 1881-1886 & \textbf{0.292 (0.027)} & \textbf{0.418 (0.141)} & \textbf{%
0.349 (1.017)} & \textbf{0.479 (0.023)} & \textbf{0.636 (0.090)} & \textbf{%
0.752 (0.494)} & \multicolumn{1}{l}{\textbf{0.976 (0.013)}} &
\multicolumn{1}{l}{\textbf{0.979 (0.029)}} & \multicolumn{1}{l}{\textbf{%
0.993 (0.190)}} \\
USA 1886-1894 & 0 (0.049) & 0 (1.628) & 0 (9.958) & \textbf{0.792 (0.014)} &
\textbf{0.760 (0.069)} & \textbf{0.759 (0.488)} & \multicolumn{1}{l}{\textbf{%
0.920 (0.012)}} & \multicolumn{1}{l}{\textbf{0.993 (0.023)}} &
\multicolumn{1}{l}{\textbf{0.996 (0.173)}} \\
USA 1953-1977 & 0 (0.024) & 0 (5.439) & 0 (35.790) & 0.011 (0.008) & \textbf{%
0.060 (0.431)} & 0.030 (2.923) & \multicolumn{1}{l}{0.025 (0.007)} &
\multicolumn{1}{l}{\textbf{0.175 (0.261)}} & \multicolumn{1}{l}{\textbf{%
0.139 (1.680)}} \\
&  &  &  &  &  &  &  &  &  \\
& Pareto &  &  & LNt &  &  &  &  &  \\
& KS & CM & AD & KS & CM & AD &  &  &  \\
Canada 1901-1916 & \textbf{0.965 (0.032)} & \textbf{0.960 (0.034)} & \textbf{%
0.420 (0.891)} & \textbf{0.925 (0.035)} & \textbf{0.948 (0.037)} & \textbf{%
0.398 (0.928)} &  &  &  \\
Canada 1945-1975 & \textbf{0.865 (0.025)} & \textbf{0.698 (0.079)} & \textbf{%
0.582 (0.673)} & \textbf{0.906 (0.024)} & \textbf{0.905 (0.045)} & \textbf{%
0.769 (0.479)} &  &  &  \\
Netherlands 1890-1935 & \textbf{0.285 (0.029)} & \textbf{0.197 (0.243)} &
\textbf{0.084 (2.072)} & \textbf{0.423 (0.026)} & \textbf{0.570 (0.103)} &
\textbf{0.166 (1.547)} &  &  &  \\
Netherlands 1946-1975 & \textbf{0.821 (0.031)} & \textbf{0.692 (0.080)} &
\textbf{0.104 (1.898)} & \textbf{0.593 (0.038)} & \textbf{0.535 (0.111)} &
\textbf{0.083 (2.076)} &  &  &  \\
France 1890-1935 & \textbf{0.997 (0.021)} & \textbf{0.986 (0.027)} & \textbf{%
0.958 (0.271)} & \textbf{0.866 (0.031)} & \textbf{0.965 (0.033)} & \textbf{%
0.946 (0.289)} &  &  &  \\
USA 1881-1886 & \textbf{0.420 (0.063)} & \textbf{0.393 (0.149)} & \textbf{%
0.396 (0.931)} & \textbf{0.560 (0.056)} & \textbf{0.782 (0.065)} & \textbf{%
0.858 (0.390)} &  &  &  \\
USA 1886-1894 & \textbf{0.163 (0.044)} & \textbf{0.118 (0.321)} & 0.042
(2.629) & \textbf{0.924 (0.022)} & \textbf{0.915 (0.043)} & \textbf{0.256
(1.230)} &  &  &  \\
USA 1953-1977 & \textbf{0.404 (0.026)} & \textbf{0.244 (0.213)} & \textbf{%
0.206 (1.387)} & \textbf{0.872 (0.018)} & \textbf{0.962 (0.034)} & \textbf{%
0.862 (0.387)} &  &  &  \\
\bottomrule &  &  &  &  &  &  &  &  &
\end{tabular}%
}
\caption{Outcomes of the Kolmogorov--Smirnov (KS), Cram\'er--von Mises (CM)
and Anderson--Darling (AD) tests, in the format $p$-value (test statistic).
Non-rejections at the 5\% level are marked in bold. Upper panels for the
whole samples, lower panel for the truncated samples.}
\label{KSCMAD}
\end{table}

We also consider a more formal approach to goodness-of-fit with standard
statistical tests. Table~\ref{KSCMAD} presents the $p$-values (and test
statistics in parentheses)\ for the Kolmogorov--Smirnov (KS), Cram\'{e}%
r--von Mises (CM) and Anderson--Darling (AD) tests.\footnote{%
Based on Montecarlo simulation using 350 synthetic data sets for each test,
sample and distribution.} The AD test is very well suited to studying the
fit in the tails \citep{Cir13}. The $p$-values for all goodness-of-fit tests
that we consider in Table~\ref{KSCMAD} indicate that the STEXP distribution
is soundly rejected as a plausible fit to the data for all the countries and
times we consider. The $p$-values in Table~\ref{KSCMAD} also show that the
2LN and 3LN distributions are very appropriate when considering the full
range of the lost person days measure, and both the Pareto and LNt describe
well the upper tail of the distributions. However, we should note that the
power of these tests increases with the number of observations %
\citep{RazWah2011}, so it might the case that the non-rejection of the
different models is favored for the truncated samples.

\begin{table}[htbp]
\centering {\tiny
\begin{tabular}{lrrrrrrrrrrrr}
\toprule & \multicolumn{1}{l}{STEXP} &  &  &  &  &  &  &  &  &  &  &  \\
& \multicolumn{1}{l}{log-likelihood} & \multicolumn{1}{l}{AIC} &
\multicolumn{1}{l}{BIC} & \multicolumn{1}{l}{HQC} &  &  &  &  &  &  &  &  \\
Canada 1901-1916 & -10883.8 & 21771.5 & 21781.8 & 21775.4 &  &  &  &  &  &
&  &  \\
Canada 1945-1975 & -102217 & 204439 & 204453 & 204444 &  &  &  &  &  &  &  &
\\
Netherlands 1890-1935 & -55040.9 & 110086 & 110100 & 110091 &  &  &  &  &  &
&  &  \\
Netherlands 1946-1975 & -14111.7 & 28227.3 & 28238.5 & 28231.4 &  &  &  &  &
&  &  &  \\
France 1890-1935 & -134924 & 269851 & 269867 & 269856 &  &  &  &  &  &  &  &
\\
USA 1881-1886 & -11660.3 & 23324.5 & 23334.8 & 23328.4 &  &  &  &  &  &  &
&  \\
USA 1886-1894 & -18821.6 & 37647.2 & 37658.7 & 37651.4 &  &  &  &  &  &  &
&  \\
USA 1953-1977 & -548415 & 1096833 & 1096852 & 1096839 &  &  &  &  &  &  &  &
\\
&  &  &  &  &  &  &  &  &  &  &  &  \\
& \multicolumn{1}{l}{LN} &  &  &  & \multicolumn{1}{l}{2LN} &  &  &  &
\multicolumn{1}{l}{3LN} &  &  &  \\
& \multicolumn{1}{l}{log-likelihood} & \multicolumn{1}{l}{AIC} &
\multicolumn{1}{l}{BIC} & \multicolumn{1}{l}{HQC} & \multicolumn{1}{l}{
log-likelihood} & \multicolumn{1}{l}{AIC} & \multicolumn{1}{l}{BIC} &
\multicolumn{1}{l}{HQC} & \multicolumn{1}{l}{log-likelihood} &
\multicolumn{1}{l}{AIC} & \multicolumn{1}{l}{BIC} & \multicolumn{1}{l}{HQC}
\\
Canada 1901-1916 & -10725.4 & 21454.9 & \textbf{21465.1} & 21458.7 & -10718.9
& 21447.77 & 21473.38 & 21457.4 & -10709.6 & \textbf{21435.24} & 21476.21 &
\textbf{21450.65} \\
Canada 1945-1975 & -100935 & 201874 & 201889 & 201879 & -100847.8 & 201705.68
& 201742.42 & 201718.03 & -100803 & \textbf{201622.76} & \textbf{201681.54}
& \textbf{201642.52} \\
Netherlands 1890-1935 & -54104.6 & 108213 & 108227 & 108218 & -54053.05 &
108116.1 & \textbf{108150.59} & \textbf{108127.96} & -54047 & \textbf{%
108110.04} & 108165.23 & 108129.02 \\
Netherlands 1946-1975 & -13717.3 & 27438.6 & 27449.8 & 27442.7 & -13647.72 &
27305.43 & \textbf{27333.45} & \textbf{27315.72} & -13643.3 & \textbf{%
27302.65} & 27347.49 & 27319.11 \\
France 1890-1935 & -132433 & 264871 & 264886 & 264876 & -132220.89 &
264451.78 & 264490.26 & 264464.5 & -132199 & \textbf{264413.7} & \textbf{%
264475.27} & \textbf{264434.05} \\
USA 1881-1886 & -11529.7 & 23063.4 & \textbf{23073.7} & \textbf{23067.3} &
-11523.89 & 23057.78 & 23083.63 & 23067.48 & -11519.6 & \textbf{23055.21} &
23096.57 & 23070.73 \\
USA 1886-1894 & -18448.4 & 36900.8 & 36912.3 & 36905 & -18388.15 & 36786.31
& \textbf{36815.03} & \textbf{36796.78} & -18383 & \textbf{36781.99} &
36827.95 & 36798.74 \\
USA 1953-1977 & -540534 & 1081072 & 1081091 & 1081078 & -540116.21 & 1080242
& 1080288 & 1080256 & -540024 & \textbf{1080064} & \textbf{1080136} &
\textbf{1080086} \\
&  &  &  &  &  &  &  &  &  &  &  &  \\
& \multicolumn{1}{l}{Pareto} &  &  &  & \multicolumn{1}{l}{LNt} &  &  &  &
&  &  &  \\
& \multicolumn{1}{l}{log-likelihood} & \multicolumn{1}{l}{AIC} &
\multicolumn{1}{l}{BIC} & \multicolumn{1}{l}{HQC} & \multicolumn{1}{l}{
log-likelihood} & \multicolumn{1}{l}{AIC} & \multicolumn{1}{l}{BIC} &
\multicolumn{1}{l}{HQC} &  &  &  &  \\
Canada 1901-1916 & -2604.81 & \textbf{5211.61} & \textbf{5215.12} & \textbf{%
5213.02} & -2608.22 & 5220.43 & 5227.45 & 5223.26 &  &  &  &  \\
Canada 1945-1975 & -6891.99 & \textbf{13785.98} & \textbf{13790.32} &
\textbf{13787.68} & -6891.58 & 13787.2 & 13795.8 & 13790.5 &  &  &  &  \\
Netherlands 1890-1935 & -11484.5 & \textbf{22970.98} & \textbf{22976.02} &
\textbf{22972.88} & -11488.9 & 22981.9 & 22991.9 & 22985.7 &  &  &  &  \\
Netherlands 1946-1975 & -3826.84 & \textbf{7655.68} & \textbf{7659.73} &
\textbf{7657.28} & -3833.46 & 7670.92 & 7679.01 & 7674.12 &  &  &  &  \\
France 1890-1935 & -4618.58 & \textbf{9239.15} & \textbf{9243.08} & \textbf{%
9240.71} & -4619 & 9242 & 9249.85 & 9245.12 &  &  &  &  \\
USA 1881-1886 & -2205.59 & 4413.18 & \textbf{4416.47} & 4414.51 & -2203.08 &
\textbf{4410.17} & 4416.73 & \textbf{4412.83} &  &  &  &  \\
USA 1886-1894 & -6419.3 & \textbf{12840.61} & \textbf{12845.07} & \textbf{%
12842.34} & -6419.67 & 12843.3 & 12852.3 & 12846.8 &  &  &  &  \\
USA 1953-1977 & -14405.3 & \textbf{28812.59} & \textbf{28817.64} & \textbf{%
28814.49} & -14405.5 & 28815 & 28825.1 & 28818.8 &  &  &  &  \\
\bottomrule &  &  &  &  &  &  &  &  &  &  &  &
\end{tabular}
}
\caption{Maximum log-likelihoods, and outcomes of the AIC, BIC and HQC
information criteria. The minimum values of these criteria for each sample
are marked in bold. Upper panels for the whole samples, lower panel for the
truncated samples.}
\label{llAICBICHQC}
\end{table}

In order to choose among the hypothesized models, we have computed the
Akaike Information Criterion (AIC), the Ba\-ye\-sian or Schwarz Information
Criterion (BIC) and the Hannan--Quinn Information Criterion (HQC), see Table~%
\ref{llAICBICHQC}. We note that the STEXP distribution is never chosen for
any sample. The lognormal distribution is chosen by the BIC for the sample
of Canada 1901-1916 and by both BIC and HQC for the sample of USA 1881-1886,
so in these cases a single lognormal could provide a good description to the
data. But in general, the information criteria differ in terms of the best
distribution as they provide some conflicting evidence. The AIC tends to
favor the 3LN, i.e., the mixture of 3 lognormal distributions. The BIC and
HQC, often select the 3LN, but also favor the lognormal and 2LN, i.e., the
mixture of 2 lognormal distributions. However, in the case of conflicting
cases \cite{BurAnd02,BurAnd04} recommend relying on the AIC, based on
information-theoretic arguments.

For the upper tail samples, it is feasible to compare the Pareto with the
LNt in another way: since they are non-nested models, it is possible to
perform Vuong tests \citep{Vuo89} to assess whether the models are
statistically equivalent (null hypothesis) or one of the models is favored
(alternative hypothesis). Positive test statistics are supportive of the
power law distribution, but negative values would suggest the truncated
lognormal distribution would be more appropriate. We see in Table~\ref{Vuong}
that we cannot distinguish between the Pareto and the LNt distributions as
are statistically equivalent models most of the cases. The two exceptions
are Canada 1901-1916 and Netherlands 1946-1935, where the tests support a
power law distribution.

\begin{table}[htbp]
\centering {\tiny
\begin{tabular}{ll}
\toprule & Vuong Pareto vs LNt \\
Canada 1901-1916 & 0.034 (2.125) \\
Canada 1945-1975 & \textbf{0.849 (-0.191)} \\
Netherlands 1890-1935 & \textbf{0.244 (1.166)} \\
Netherlands 1946-1975 & 0.005 (2.784) \\
France 1890-1935 & \textbf{0.652 (0.451)} \\
USA 1881-1886 & \textbf{0.265 (-1.114)} \\
USA 1886-1894 & \textbf{0.931 (0.087)} \\
USA 1953-1977 & \textbf{0.945 (0.069)} \\
\bottomrule &
\end{tabular}%
}
\caption{Outcomes of Vuong tests of Pareto vs LNt distributions for the
upper tail. The format is $p$-value (test statistic). Non-rejections at the
5\% level are marked in bold.}
\label{Vuong}
\end{table}

{}From a statistical perspective, the lognormal distribution is very often
inferior to a mixture of lognormal distributions. This is not surprising as
the summary statistics indicate that there is a great deal of skew and
kurtosis in the lost person days measure of strike size and a mixture of
lognormal distributions is better able to capture this feature of the data.
The AIC, the preferred information criterion, suggests that this is true
across the countries and time periods we consider. This consistency of our
finding suggests that there could a common process shaping strikes, e.g.,
legislative, economic or social factors, across time and the countries we
consider. In the upper tail of the data, it is much more difficult to
distinguish the power law from the truncated lognormal distribution, but
this should not be overly surprising given the results in recent empirical
work \citep{Per05,BeeRicSch11,BeeRicSch13,BeeRicSch19,SchTre18}.

\section{Conclusions}

\label{discussionconcl}

We have studied the distribution of strike size using data from the
late-19th and 20th centuries in some European countries, Canada and the US.
There are remarkable regularities for this variable that reflect in its
parametric distribution. First, we find that strike size, for the different
samples we consider, can be faithfully described by a mixture of two or
three lognormals (2LN, 3LN in the notation above), as it happens as well for
some samples of city sizes %
\citep{KwoNad19,BanChiPrePueRam19,Su19,PueRamSan20}. Second, finding a
common distribution across countries and time periods also suggests that the
distribution of strikes could be shaped by common forces, which could be
economic, legislative or social in nature. Third, when restricting the data
to the upper tail, the identification of the appropriate model becomes more
problematic. We have studied, because of their importance, the Pareto and
(upper-tail) truncated lognormal. The results of Vuong test show that they
are statistically equivalent in most of the cases we consider, although
there are some exceptions that favor the Pareto. Thus the upper tail can be
described by a power law, although it should not considered as a unique
model for this purpose.

\section*{Author contributions}

Michele Campolieti: Conceptualization, data curation, formal analysis,
investigation, methodology, software, supervision, validation,
visualization, writing-original draft, writing-review \& editing. Arturo
Ramos: Conceptualization, data curation, formal analysis, funding
acquisition, investigation, methodology, resources, software, validation,
visualization, writing-original draft, writing-review \& editing.

\section*{Competing interests statement}

The authors declare to have no competing interests concerning the research
carried out in this article.

\section*{Acknowledgments}

The work of Arturo Ramos has been supported by the Spanish \emph{Ministerio
de Econom\'{\i}a y Competitividad} (ECO2017-82246-P) and by Aragon
Government (ADETRE Reference Group).

\section*{Data appendix}

The data for the Dutch strikes were obtained from the following webpage:

\href{https://collab.iisg.nl/web/labourconflicts/datafiles} {%
https://collab.iisg.nl/web/labourconflicts/datafiles}

The Dutch data, known as SIN (Strikes in the Netherlands or, equivalently,
Stakingen in Nederland) were collected by Sjaak van der Velder and are
described in \cite{Vel03}.

French strike data were collected by \cite{ShoTil74} and are available via
the ICPSR at the University of Michigan.

\href{https://www.icpsr.umich.edu/index.html}{https://www.icpsr.umich.edu/index.html}

The Canadian data for 1901-1916 were obtained via the Labour Conflicts
Dataverse

\href{https://datasets.socialhistory.org/dataset.xhtml?persistentId=hdl:10622/J5TJNF}{https://datasets.socialhistory.org/dataset.xhtml?persistentId=hdl:10622/J5TJNF}

They were compiled by \cite{Ros17} based on the records kept by the
Department of Labour, Canada.

The Canada data for 1946-1975 were obtained from the Labour Program,
Government of Canada by request

\href{https://www.canada.ca/en/employment-social-development/services/labour-contact.html}{https://www.canada.ca/en/employment-social-development/services/labour-contact.html}

U.S. data for 1881-1886 and 1886-1894 were collected from the 3rd and 10th
Annual Reports of the U.S. Commissioner of Labour \citep{us1888,us1896}.

U.S. data for 1953-1977 are available via the ICPSR at the University of
Michigan.

\bibliographystyle{apalike}
\bibliography{biblio}




\end{document}